\documentstyle[aps]{revtex}

\newfont{\bldm}{cmmib10}

\def\delbol{\mbox{\bldm\symbol{"0E}}}

\def\xibol{\mbox{\bldm\symbol{"18}}}
\def\pibol{\mbox{\bldm\symbol{"19}}}

\def\phibol{\mbox{\bldm\symbol{"1E}}}

\def\half{{\textstyle{1\over 2}}}
\def\Tr{{\rm Tr}}

\def\qbar{{\overline q}}

\def\A{{\scriptscriptstyle A}}
\def\V{{\scriptscriptstyle V}}
\def\L{{\scriptscriptstyle L}}
\def\R{{\scriptscriptstyle R}}
\def\NP#1{Nucl.\ Phys.\ {\bf  #1}}         
\def\PR#1{Phys.\ Rev.\ {\bf  #1}}
\def\PRL#1{Phys.\ Rev.\ Lett.\ {\bf  #1}}
\def\PL#1{Phys.\ Lett.\ {\bf  #1}}

\def\group#1{${\rm SU}(#1)_\L\times{\rm SU}(#1)_\R$}
\def\su3v{${\rm SU}(3)_\V$}
\def\nf{N_f}
\def\nfbar{\overline{N}\!{}_f}
\def\ua1{U(1)${}_\A$}

\begin{document}

\title{\ua1 Symmetry and Correlation Functions in the High Temperature 
Phase of QCD}

\author{Michael C. Birse$^{a,b}$, Thomas D. Cohen$^a$ and Judith A.
McGovern$^{a,b}$}

\address{$^a$ Department of Physics,\\
University of Maryland, College Park, MD, 20742-4111,USA\\ \ }

\address{$^b$ Theoretical Physics Group, Department of Physics and Astronomy,\\
University of Manchester, Manchester M13 9PL, U.K.}
\date{\today}
\maketitle

\begin{abstract}
Simple group-theoretical arguments are used to demonstrate that in the high
temperature (chirally restored) phase of  QCD with $\nf$ massless flavours, all 
$n$-point correlation functions of quark bilinears  are invariant under \ua1
transformations provided $n < N_f$.  In particular this implies that the
two-point correlation function in the $\eta'$ channel is  identical to that in
the pion channel for $\nf >2$.  Unlike previous work, this result does not
depend on the topological properties of QCD and can be formulated without
explicit reference to functional integrals. \vspace{.5in} \end{abstract}

The nature of the high temperature phase of QCD is a problem of considerable
importance. The approximate \group{\nf} chiral symmetry of QCD in this  phase is
believed to be broken only by non-zero quark masses. The study of this phase
is both intrinsically interesting and of relevance to  cosmology and
ultra-relativistic heavy ion collisions.  One aspect of the problem that has
received  recent interest 
\cite{shuryak,christ,cohen,hatsuda,evans} is the role of the 
anomalously broken \ua1 symmetry in the chirally restored phase.

 From a theoretical perspective,
the natural way to study the possible role of \ua1 symmetry breaking  in
the high temperature phase of QCD is to study correlation functions of composite
operators constructed from quark and gluon fields and to compare
multi-point correlation functions that are related  by \ua1 symmetry (and
perhaps also \group{\nf} symmetry).  All observable manifestations of anomalous
\ua1 symmetry breaking should be reflected in the behaviour of these
correlation functions.  It has  been suggested by Shuryak that the \ua1
symmetry might be restored along with \group{\nf} in the high temperature
phase, in the sense that no \ua1-violating effects can be found among
correlation functions in this phase \cite{shuryak}.  Moreover, it has been
shown that unless there are contributions from configurations in the
functional integral that form a set of measure zero in the chiral limit,
\ua1-violating correlation functions must vanish \cite{cohen}.

On the other hand, Lee and Hatsuda (LH) \cite{hatsuda} and Evans, Hsu and
Schwetz (EHS) \cite{evans} have argued that contributions from the 
winding-number-one sector do not vanish and are exactly the kind of zero-measure
contributions that exploit the loophole in ref.\cite{cohen}. From studying the
form of the functional determinant, LH and EHS conclude that for QCD with two
light flavours the effects of \ua1 violation can be seen in the study of
two-point functions of quark bilinears.  However, for three or more light
flavours they conclude that \ua1 violation cannot be observed in two-point
functions.  For example, disregarding explicit symmetry breaking due to the
quark masses, the correlation functions in the $\pi$, $\sigma$ and $\eta'$
channels are identical.  This in turn means that the screening masses $m_\pi$,
$m_\sigma$ and $m_{\eta'}$ are all degenerate.  This may seem startling at
first glance, since it has been  well known since 't Hooft's seminal papers
\cite{'t Hooft} that the anomaly together with topology solves the \ua1
problem, allowing the $\eta'$ to be split from the pion.  If LH and EHS are
correct, however, for $\nf \ge 3$ the splitting of the $\pi$ from the $\eta'$
for $m_q=0$ depends on  spontaneous breaking of the chiral symmetry. They argue
more generally that $n$-point correlations of quark bilinears will have no
\ua1-violating effects for $n<\nf$ but can, and in general will, have
observable \ua1 violation for $n \ge \nf$.

The arguments of LH and EHS are highly suggestive but not definitive.  In
particular, the argument depends on taking the infinite-volume limit after the
chiral limit, which might not be permissible since issues
associated with symmetry breaking often depend on taking the infinite-volume
limit first.  However, as we  will show in this letter,
the conclusion of EHS that all \ua1-violating $n$-point correlation functions
of quark bilinears vanish for $n < \nf$ is, in fact, correct.  This includes
the somewhat counter-intuitive result that the $\pi$ and $\eta'$ channels are
degenerate above $T_c$ for three or more flavours.  The proof given here is
quite simple and relies only on basic group theory; it does not depend
explicitly on topological properties of QCD.

Before addressing the problem in its general form, we will turn our attention
to a very simple demonstration that in the chiral limit the $\pi$
and $\eta'$ two-point correlation functions are degenerate above $T_c$ for
three flavour QCD.
For simplicity of notation, we denote pseudoscalar quark-bilinear
composite fields as $\phi_a$ and scalar quark bilinears as 
$\xi_a$, with subscripts $a=0$ for the singlets and $a=1,\ldots \nf^2-1 $ 
for the rest (such as the pion triplet in SU(2) or the octet in SU(3)). Thus 
for example the quark bilinear $\qbar(x) i \gamma_5 \lambda_a q(x)$ is denoted
$\phi_a(x)$,
where in addition to the $\nf^2-1$ generalised Gell-Mann matrices we have 
introduced $\lambda_0$, defined as $\sqrt{2/\nf}$ times the unit matrix.

Suppose, taking $\nf=3$, we consider a two-point correlation function in a 
pionic channel---for example the $\phi_3$  correlator:
\begin{equation}
 \langle \qbar (x)i\gamma_5 \lambda_3 q(x)\; \qbar(0)i\gamma_5\lambda_3 q(0)
\rangle \equiv \langle \phi_3(x) \, \phi_3(0) \rangle \; \;   .
\label{picor}
\end{equation}
Now consider what happens to this correlator under a 
particular ${\rm SU}(3)$ axial transformation
\begin{equation}
q \, \rightarrow \, q' \, = \,e^{ i\gamma_5 \left ( \sqrt{3} \lambda_8 -
  \lambda_3 \right) \frac{\pi}{4} } \, q\; \;   .
 \label{chirot}
 \end{equation}
 It is simple to show that under this transformation
 $\phi_3$ transforms in the following way
\begin{equation} 
 \phi_3 \, \rightarrow \, \phi^\prime_3 \, =\, \sqrt{\frac{2}{3}} \phi_0 + 
 \sqrt{\frac{1}{3}} \phi_8\; \;   .
\label{pirot}
 \end{equation}
 Thus, under the axial rotation in eq.~(\ref{chirot}) the
 correlation function in eq.~(\ref{picor}) transforms according
 to
\begin{eqnarray}
\hspace*{-.45in} \langle \phi_3(x) \phi_3(0) \rangle  \, \rightarrow \,
 \langle  \phi^\prime_3(x) \phi^\prime_3(0) \rangle & = & \,
 \frac{2}{3}  \langle \phi_0(x) \phi_0(0) \rangle \, + \,
  \frac{1}{3} \langle \phi_8(x) \phi_8(0) \rangle \, \nonumber \\ 
&& + \, \frac{\sqrt{2}}{3} \langle  \phi_0(x) \phi_8(0)
 \rangle \, + \,\frac{\sqrt{2}}{3} \langle  \phi_8(x) \phi_0(0)\rangle\; \; . 
\label{cortran}
\end{eqnarray}
 
Neglecting explicit \su3v breaking due to quark masses,  all thermal
expectation values must be \su3v singlets since flavour symmetry is not
spontaneously broken. This means that the last two terms of
eq.~(\ref{cortran}) are  identically zero. Moreover \su3v symmetry implies that
the $\phi_3$ correlator  must be equal to $\phi_8$ correlator.  Thus,
eq.~(\ref{cortran}) can be  rewritten as
\begin{equation}
   \langle  \phi^\prime_3(x) \phi^\prime_3(0) \rangle \, =
 \frac{2}{3}  \langle \phi_0(x) \phi_0(0) \rangle  \, +
 \, \frac{1}{3} \langle \phi_3(x) \phi_3(0) \rangle\; \;   .
 \label{equiv}
 \end{equation}    If one is studying the chirally restored phase, as we  are
here, then by definition all correlation functions  are invariant under
arbitrary \group{3}  transformations, including eq.~(\ref{chirot}).   Thus
$\langle  \phi_3(x) \phi_3(0) \rangle = \langle \phi^\prime_3(x) 
\phi^\prime_3 (0) \rangle$, which along with  eq.~(\ref{equiv}) implies that
$\langle  \phi_3(x) \phi_3(0) \rangle = \langle \phi_0(x)  \phi_0 (0)
\rangle$. In other words the two-point  correlation function in the pion
channel is identical to the  correlation function in the  $\eta'$ channel. 
This completes the demonstration. For more than three flavours, it is easy to
generalise the preceding argument, with the same result.
However, it should be noted that the
argument does not work for two flavours, since there is  no analogue of
the $\lambda_8$ rotation which is essential to obtain eq.~(\ref{pirot}).

Now let us consider the problem more generally. We will use the fact that
only \group{\nf}-symmetric multipoint correlation functions have a 
thermal expectation value in the restored phase to restrict severely the 
number of possible independent $n$-point functions for low $n$, and show
that for $n<\nf$ they are all invariant under \ua1 transformations.

The quark bilinears transform under \group{\nf} as the direct sum of 
$(\nf,\nfbar)$ and $(\nfbar,\nf)$, as can be seen by writing, for instance,
$\phi_3\equiv\qbar i\gamma_5\lambda_3 q$ as $i(q_\L^\dagger\gamma_0\lambda_3 
q_\R - q_\R^\dagger\gamma_0 \lambda_3 q_\L)$.  From them we can form linear 
combinations that have definite transformation properties, namely
\begin{eqnarray}
&&{\bf M}=\sum_{a=0}^{\nf^2-1} (\xi_a+i\phi_a)\lambda_a\quad:\quad 
(\nf,\nfbar)\\
&&{\bf M}^\dagger=\sum_{a=0}^{\nf^2-1} (\xi_a-i\phi_a)\lambda_a\quad:\quad
(\nfbar,\nf)
\end{eqnarray}
Under \group{\nf} transformations $\bf M$ and $\bf M^\dagger$ transform
as ${\bf M\to U}_\L{\bf M U}_\R^\dagger $ and ${\bf M^\dagger\to U}_\R
{\bf M^\dagger U}_\L^\dagger$. We will also need the parity transformation, 
${\bf M}\to {\bf M}^\dagger$, and the behaviour under \ua1 rotations through
$\theta$, ${\bf M}\to e^{i\theta}{\bf M }$.

 There are two differences between $\nf=2$ and $\nf >2$: 
firstly, $\overline{2}\equiv 2$ so that $\bf M$ and $\bf M^\dagger$ have the
same tensor structure, and secondly the symmetric structure constants
$d_{abc}$ vanish for SU(2).  As a result there are two independent chiral
multiplets $(\xi_0,\phibol)$ and $(\phi_0,\xibol)$ (better known as
($\sigma$,$\pibol$) and ($\eta'$, $\delbol$)\,).  However for $\nf>2$, no such
separation occurs and all $2\nf^2$ mesons transform into one another under a
general chiral transformation.

In the restored phase of \group{\nf}, the only vacuum correlators which can 
be non-vanishing are \group{\nf} singlets (``chiral singlets") that are also
even under parity.  (We are neglecting effects due to finite current-quark
masses which explicitly break the symmetry.)  Thus none of the bilinears have
vacuum expectation values---for instance, $\langle\qbar(x)q(x)\rangle=0$.
Chiral singlets can be constructed in two distinct ways.  One is to take equal
numbers of $\bf M$'s and $\bf M^\dagger$'s, coupled up to a singlet.  Examples
are $\Tr\,({\bf M}_1^\dagger {\bf M}_2)$, $\Tr\,({\bf M}_1^\dagger {\bf M}_2
{\bf M}_3^\dagger {\bf M}_4)$, $\Tr\,({\bf M}_1^\dagger {\bf M}_2) \Tr\,({\bf
M}_3^\dagger {\bf M}_4)$, etc.\ (where ${\bf M}_i \equiv {\bf M}(x_i)$).
Only even-parity combinations have vacuum expectation values, giving one
independent two-point function $\Tr\,({\bf M}_1^\dagger {\bf M}_2)+\Tr\,({\bf
M}_2^\dagger {\bf M}_1)$, twelve independent four-point functions (six of
which involve products of two-point functions) and so on.  All of these are
not only chiral invariants; they are obviously \ua1-invariant as well.

The other way of obtaining a singlet is to couple $\nf$ $\bf M$'s or $\nf$ $\bf 
M^\dagger$'s together. This produces  two singlets, 
\begin{equation}
{1\over (\nf) !}\epsilon_{ijk\ldots p} \epsilon_{i'j'k'\ldots p'} 
({\bf M}_1)_{ii'}({\bf M}_2)_{jj'}({\bf M}_3)_{kk'}\ldots
({\bf M}_{\scriptscriptstyle\nf})_{pp'}
\label{det}
\end{equation}
and the analogous expression with $\bf M\to M^\dagger$. (All indices run from 
1 to $\nf$.)  For identical $\bf M$'s, these 
terms are just $\det {\bf M}$ and $\det {\bf M^\dagger}$.  By parity, only the 
sum has a non-vanishing vacuum expectation value.  This, however, is {\it not}
\ua1-invariant.  Further chiral singlet, \ua1-violating terms
may be obtained by coupling, for instance, $\nf+1$ $\bf M$'s and an $\bf 
M^\dagger$, etc.

The crucial point is that \ua1-violating terms can only be obtained from 
structures such as eq.~(\ref{det}) involving at least $\nf$ bilinears,
so for $n<\nf$ all $n$-point
functions in the chirally restored phase are \ua1-invariant, completely
independently of the strength of the  anomaly.  

For $\nf=2$, there are two chiral-singlet two-point functions with even parity,
 one \ua1-invariant and one \ua1-violating.  If the non-perturbative effects
of the anomaly persist
above the chiral restoration point, the $\eta'-\eta'$ and $\pi-\pi$ 
correlators will be different.  However for $\nf=3$ or higher, there is only
one, \ua1-invariant, chiral singlet, and the $\eta'-\eta'$ and 
$\pi-\pi$ correlators must be equal.  In other words, the $\eta'$ and pion will
be degenerate, as was already shown above.

The above arguments are easily extended to more general quark bilinears.
Vector and axial-vector bilinears, $\qbar \gamma_\mu\lambda_a q$ and
$\qbar i\gamma_\mu\gamma_5\lambda_a q$, are themselves \ua1 singlets, and so
their correlators can never violate \ua1. The tensor bilinears can be grouped
into three sets, ($\qbar \sigma^{{\scriptscriptstyle 0}i}\lambda_a q$, 
$\half \epsilon_{ijk}\qbar \sigma^{jk}\lambda_a q$), for $i=1-3$. For each $i$,
these transform under \group{\nf} and \ua1 exactly like the scalars and
pseudoscalars ($\xi_a$, $\phi_a$).   Thus the arguments above can be
repeated exactly, with the same conclusion: the lowest \ua1-violating
correlators are $\nf$-point functions.

The notation we have used is that of the $\nf$-flavour linear sigma
model \cite{lee}, and of course similar arguments have long been used in
constructing the $\nf$ invariants of that model, which are $\Tr[({\bf
M^\dagger M})^n]$, $n<\nf$, and ($\det{\bf M}+\det{\bf M^\dagger}$). However it
must be stressed that all that the two have in common is the group structure,
and the arguments presented here are independent of any dynamical model.

These results have implications for lattice studies
of screening masses in the chirally restored phase of QCD.
Calculations showing degeneracy of the pion and delta screening masses
would give unequivocal proof of \ua1 restoration only if the symmetry group is
\group{2}.  It is not clear that any current lattice technique reproduces
this.

In summary, we have shown on purely group-theoretic grounds that, in the
chirally restored phase of \group{\nf}, \ua1 violation cannot occur in 
correlation functions of $n$ quark bilinears if $n<\nf$.  Thus for $\nf>2$, 
lattice studies of screening masses in mesonic channels cannot determine 
whether \ua1 is also restored.  This conclusion is the same as that
of refs.~\cite{hatsuda,evans}, but our argument is general and does not
make reference to the topology of the QCD vacuum.

This work was  supported in part by the US Department of Energy
under grant no. DE-FG02-93ER-40762, and in part by the UK EPSRC. M.C.B.\ and
J.McG.\ would like to thank the TQHN group at the University of Maryland for
its generous hospitality.

\end{document}